%% file: Hoefer_Intrinsic_surface_conduction_through_topological_surface_states_of_insulating_Bi2Te3_epitaxial_thin_films.tex
\documentclass[aip,
reprint,
superscriptaddress]{revtex4-1}

\usepackage{graphicx}
\usepackage{amsmath}
\usepackage[colorlinks=true,urlcolor=blue,citecolor=black]{hyperref}
\usepackage[]{natbib}

\begin{document}

\title{Intrinsic conduction through topological surface states of insulating Bi$_2$Te$_3$ epitaxial thin films}

\author{Katharina Hoefer}
\email{katharina.hoefer@cpfs.mpg.de}
\affiliation{Max Planck Institute for
Chemical Physics of Solids, N\"{o}thnitzer Strasse 40, Dresden 01187, Germany}
\author{Christoph Becker}\affiliation{Max Planck Institute for
Chemical Physics of Solids, N\"{o}thnitzer Strasse 40, Dresden 01187, Germany} \author{Diana Rata}
\affiliation{Max Planck Institute for
Chemical Physics of Solids, N\"{o}thnitzer Strasse 40, Dresden 01187, Germany}
\author{Jesse Swanson}
\affiliation{Max Planck Institute for
Chemical Physics of Solids, N\"{o}thnitzer Strasse 40, Dresden 01187, Germany}\affiliation{University of British Columbia, Vancouver,
BC, Canada V6T 1Z4}
\author{Peter Thalmeier}\affiliation{Max Planck Institute for
Chemical Physics of Solids, N\"{o}thnitzer Strasse 40, Dresden 01187, Germany}
\author{Liu Hao Tjeng}
\email{hao.tjeng@cpfs.mpg.de}
\affiliation{Max Planck Institute for
Chemical Physics of Solids, N\"{o}thnitzer Strasse 40, Dresden 01187, Germany}

\date{\today}


%\begin{article}
\begin{abstract}
Topological insulators represent a novel state of matter with surface charge carriers having a massless Dirac dispersion and locked helical spin polarization. Many exciting experiments have been proposed by theory, yet, their execution have been hampered by the extrinsic conductivity associated with the unavoidable presence of defects in Bi$_2$Te$_3$ and Bi$_2$Se$_3$ bulk single crystals as well as impurities on their surfaces. Here we present the preparation of Bi$_2$Te$_3$ thin films that are insulating in the bulk and the four-point probe measurement of the conductivity of the Dirac states on surfaces that are intrinsically clean. The total amount of charge carriers in the experiment is of order 10$^{12}$ cm$^{-2}$ only and mobilities up to 4,600 cm$^2$/Vs have been observed. These values are achieved by carrying out the preparation, structural characterization, angle-resolved and x-ray photoemission analysis, and the temperature dependent four-point probe conductivity measurement all \textit{in-situ} under ultra-high-vacuum conditions. This experimental approach opens the way to prepare devices that can exploit the intrinsic topological properties of the Dirac surface states.
\end{abstract}

\maketitle

\section{Introduction}
The observation of a quantum spin Hall effect due to edge states in 2D HgTe/CdTe quantum wells \cite{Bernevig2006,Koenig2007,Koenig2008} has opened the field of topological states in matter without external magnetic fields applied.
In 3D materials the existence of metallic surface states with massless Dirac dispersion, locked helical spin polarization and a simultaneous bulk insulating behavior is the hallmark of a topological insulator state (TI) \cite{Kane2005,Fu2007,Moore2007,Zhang2009a}.
For Bi$_2$Te$_3$ and Bi$_2$Se$_3$  \cite{Xia2009a,Chen2009,Hsieh2009} their presence and topological  protection was inferred from angle-resolved photoelectron spectroscopy (ARPES) experiments  \cite{Xia2009a,Chen2009}, Shubnikov-deHaas (SdH) oscillations and weak antilocalization in magnetotransport \cite{Ando2013} as well as quasiparticle interference with scanning tunneling microscopy (STM) \cite{Zhang2009b,Alpichshev2010}.

The Bi$_2$Te$_3$ and Bi$_2$Se$_3$ TI materials, however, have a severe problem that has so far hindered four-probe conductivity measurements involving the Dirac surface states (SS) \cite{Hsieh2009,Butch2010,Qu2010}: the presence of vacancies and anti-site defects in the bulk seems unavoidable with the result that the bulk conductivity will overwhelm the contribution of the surface states \cite{Culcer2010}. A remedy often used so far has been the application of counter doping (e.\,g. with Ca, Sn or Pb) \cite{Chen2009,Hsieh2009,Hor2009,Checkelsky2009,Aitani2013}, or gating \cite{Chen2010,Checkelsky2011,Kim2012} of exfoliated samples. Another promising way to solve the problem is to use thin films \cite{Zhang2009,Zhang2011} so that the surface to bulk ratio of the conduction is increased. By varying the growth parameters, one can change the film from p- to n-type \cite{Li2010,Wang2011}, suggesting that it should be possible to make the films consistently insulating. Another important aspect is that the conductivity studies reported so far \cite{Kim2011,Bansal2012,Taskin2012} have their samples exposed to air for mounting of the contacts before the four-probe conductivity measurements, leading again to n-doping of the samples as indicated by the shift of the Fermi level back into the bulk conduction band (BCB) \cite{Benia2011a,Chen2012}. It is conceivable that  the mobility of the surface carriers is decreased by the presence of the contaminants.

Here we followed a different route. First we identified the conditions under which molecular beam epitaxy gives stable and reproducible results for obtaining single domain and  bulk-insulating Bi$_2$Te$_3$ films. The essential point is hereby to achieve a full distillation process for the Te and to use substrates with negligible lattice mismatch. Second, the structural characterization, the spectroscopic analysis, and the conductivity measurements were done in the same ultra-high vacuum (UHV) system, ensuring that the intrinsic transport properties of the topological surface states are being recorded reliably.
We  achieved low charge carrier concentrations in the range of 2-4$\cdot10^{12}$\,cm$^{-2}$ with high mobility values up to  4,600\,cm$^2$/Vs at the surface. We also quantified the effect of air-exposure on the conductance of the films, demonstrating the necessity to protect the surface.

\section{Results and Discussion}
\subsection{Thin film growth and sample characterization}

Bi$_2$Te$_3$ films with a thickness of 10--50 quintuple layers (QLs with 1 QL $\sim$ 1\,nm)  were grown on epi-polished and vacuum-annealed BaF$_2$\,(1\,1\,1) insulating substrates. The lattice mismatch between Bi$_2$Te$_3$ and BaF$_2$ is less than 0.1\%. Elemental Bi was evaporated from an effusion cell with a rate of 1\,\AA/min and elemental Te with a rate of 8\,\AA/min. We  made use of the so-called distillation conditions \cite{Sutarto2009}, in which the excess Te is re-evaporated from  the sample to the vacuum by choosing an appropriate elevated temperature for the substrate. We  determined that 170\,$^\circ$C is the minimum temperature at which appreciable Te re-evaporation starts to occur, and that 260\,$^\circ$C is the maximum temperature at which point the Bi$_2$Te$_3$  deposits on  the substrate; Fig. S1 in Supporting Information. We found that 250\,$^\circ$C is the optimum temperature: full evaporation of excess Te, excellent surface atomic mobility for layer-by-layer growth, and a Bi (and Bi$_2$Te$_3$) sticking coefficient close to 1.  After the deposition, the Te flux is not stopped before the substrate temperature is below 210\,$^\circ$C to prevent Te deficiency at the surface.

\begin{figure}[tb]
\centerline{\includegraphics[width=.35\textwidth]{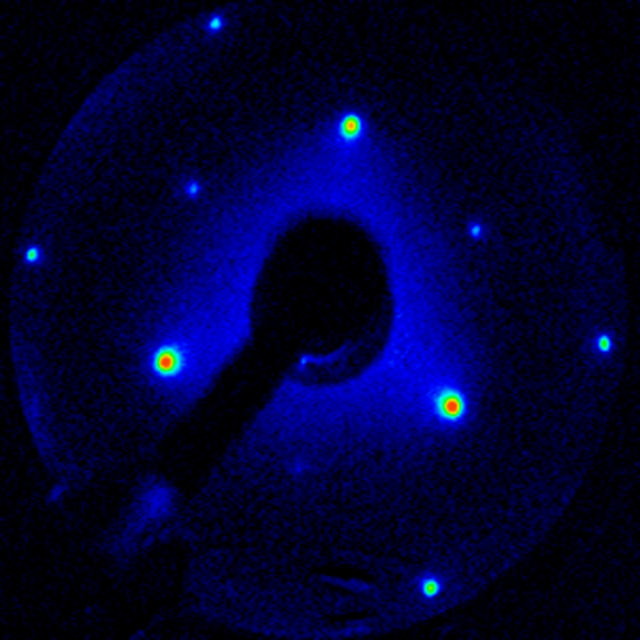}}
\caption{Bi$_2$Te$_3$ \textit{in-situ} thin film characterization. LEED image taken at 50\,eV electron energy depicts a threefold symmetry, indicating single domain growth of the films.}\label{pic1_LEED}
\end{figure}

\textit{In-situ} and real-time reflection high-energy electron diffraction (RHEED) confirms the formation of high-quality and very smooth single crystalline films, with a QL by QL growth rate of 0.3\,QL/min, consistent with the Bi supply rate; Fig. S2 in Supporting Information. Low-energy electron diffraction (LEED) shows a pattern with a threefold symmetry, demonstrating that the films are single domain (Fig.\,1). X-ray photoelectron spectroscopy (XPS) reveals very narrow and symmetric Te\,3d and Bi\,4f core level lines, indicating the absence of any Te or Bi excess; see Fig. S3.

\begin{figure}[b]
\centerline{\includegraphics[width=.4\textwidth]{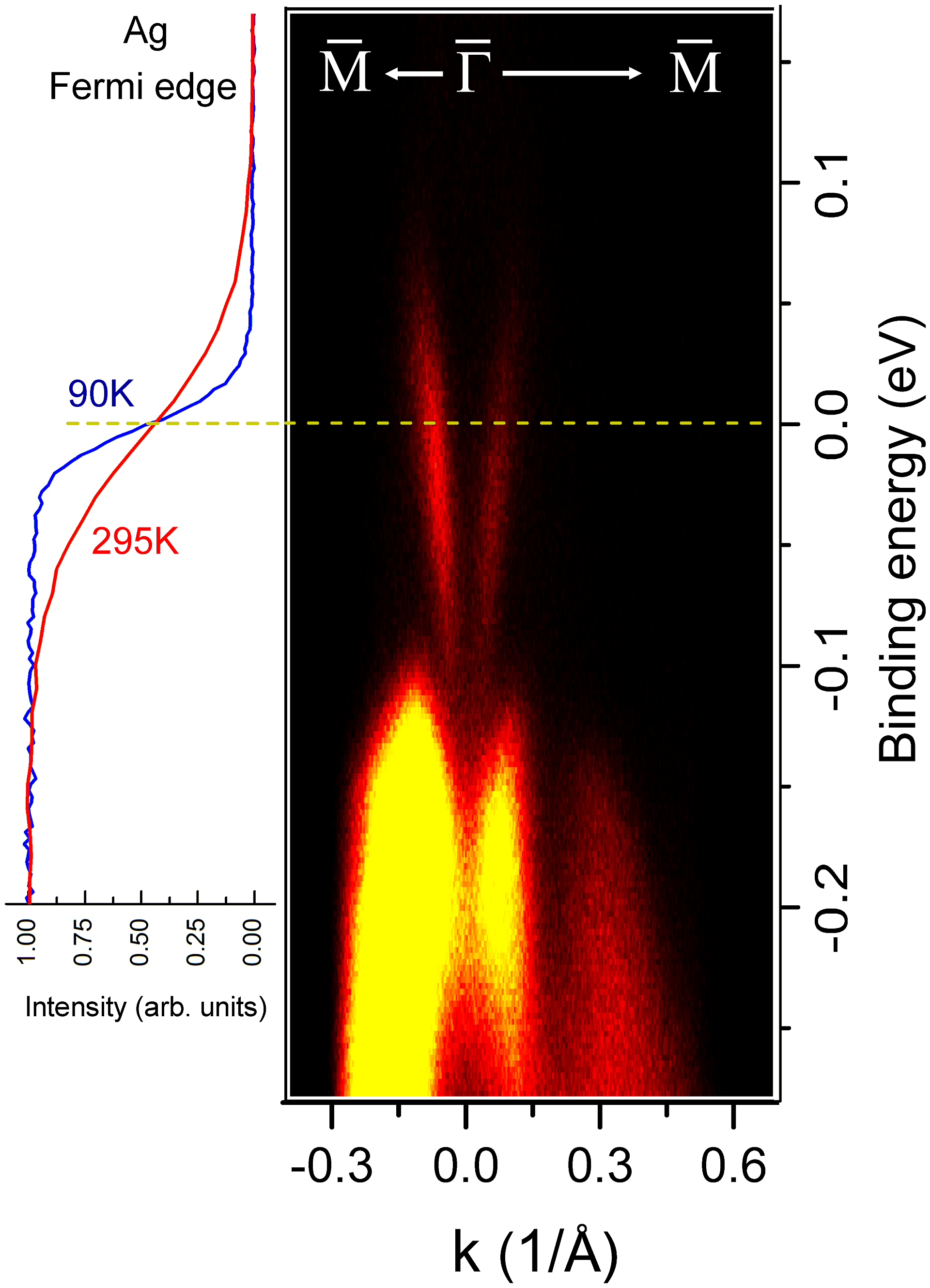}}
\caption{Valence band structure observed by angle-resolved photoemission spectroscopy. ARPES band dispersion spectrum taken along the $\bar{\Gamma}-\bar{M}$ direction of a 10\, QL thick  film at room temperature. The  dashed line indicates the position of the Fermi level, calibrated using a silver reference sample. Only the surface states intersect the Fermi level, revealing insulating bulk behavior of the films.}\label{pic2_ARPES}
\end{figure}

Figure 2 depicts the valence band ARPES in the vicinity of the Fermi level.
The measurement was carried out at room temperature on a 10\,QL thick Bi$_2$Te$_3$ film. The characteristic linear dispersion of the surface states is clearly observable, indicating the presence of the massless Dirac fermions (Dirac cone) \cite{Chen2009}.  The obtained Fermi velocities are  $\hbar v_F = (2.20\pm 0.20)$\,eV{\AA} along $\bar{\Gamma}-\bar{K}$ and  $\hbar v_F = (2.08\pm 0.20)$\,eV{\AA} along $\bar{\Gamma}-\bar M$, which agrees well with theoretically reported values of $2.13$ and $2.02$\,eV{\AA} along $\bar{\Gamma}-\bar K$ and $\bar{\Gamma}-\bar M$, respectively \cite{Zhang2009a}.  The Fermi level lies well above the bulk valence band  (BVB) maximum and well below the bulk conduction band  minimum, i.e. well within in the bulk band gap, so the conductance can be attributed only to the surface states \cite{Culcer2010}. We emphasize, that no counter doping was necessary to suppress the presence of bulk charge carriers. This situation is preferred  when studying transport phenomena of topological insulators, since counter doping introduces also disorder and decreases the surface carrier mobility \cite{Culcer2010,Zhang2011}.

Performing ARPES at room temperature rather than at low temperatures has the important advantage that also states up to $\sim$100\,meV above the Fermi level are thermally populated and therefore become visible (see also Fig. 2 for data taken from a Ag reference sample at 90 K and 295 K, showing the thermal occupation above the Fermi level). Knowing the conduction band states is important when analyzing  the conductance data. In our case, the absence of a BCB signal in the room temperature ARPES ensures that the conductance at low  temperatures is given by  the Dirac surface states.

\begin{figure*}[tb]
\begin{center}
\centerline{\includegraphics[width=0.85\textwidth]{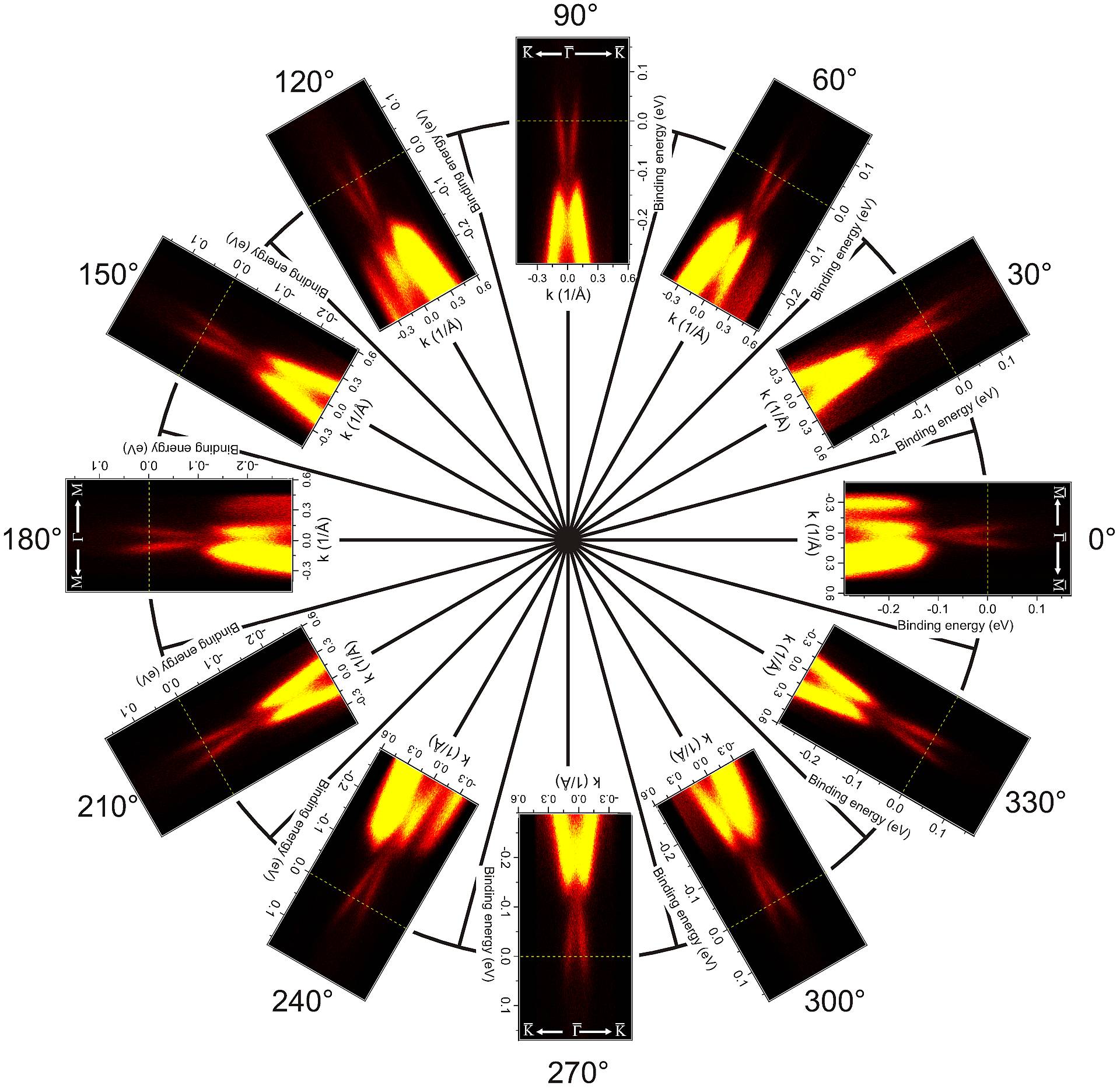}}
\caption{ARPES spectra recorded at an in-plane rotation in steps of 30$^\circ$ on a 10\,QL thick Bi$_2$Te$_3$ sample. The 3-fold symmetry can be clearly seen in the repeating valence band structure at e.g.\, 0$^\circ$, 120$^\circ$ and 240$^\circ$.}\label{pic3_360ARPES}
\end{center}
\end{figure*}

ARPES also provides spectroscopic evidence for the single domain nature of our films. Figure 3 depicts the azimuthal angular dependence of the spectra and one can clearly observe the threefold symmetry, e.g. the spectra taken at 0$^\circ$, 120$^\circ$, and 240$^\circ$ are identical and mirrored to those taken at 60$^\circ$, 180$^\circ$, and 300$^\circ$. Further characterization by \textit{ex-situ} X-ray diffraction (XRD) measurements also clearly confirm the single crystalline and single domain (threefold symmetry rather than sixfold symmetry \cite{Krumrain2011,Caha2013,Schreyeck2013}) nature of the films, see Fig. S4 and S5. In addition, from the Kiessig fringes in the X-ray reflectivity (XRR), we can deduce a long range surface roughness of less than 0.2\,nm, confirming the smoothness of our films; Fig. S6.

\subsection{In-situ transport experiments}
Special effort has been made to have the feasibility to carry out the transport measurements, in addition to ARPES, under ultra-high vacuum conditions. The importance of maintaining such conditions in order to preserve the integrity of the topological surface states was emphasized before \cite{Hirahara2010,Barreto2014}. In our set-up this was realized by a homemade non-permanent point contact mechanism, which enables \textit{in-situ} electrical contacting. Calculations show that the interface of a TI with large metallic contacts leads to changes of their properties due to hybridization  with the metallic states \cite{Culcer2010}. We use  spring-loaded point contacts to keep the influence of the metal on the topological insulator as small as possible due to the reduced contact area as compared with the usually applied \textit{ex-situ} contacting methods, like adding Ag-paint or sputtered contact pads.

The inset of Fig. 4\,A shows the voltage-current characteristics of the contacts at selected temperatures. The straight lines through zero demonstrate that the contacts are ohmic over the entire temperature range measured. There are also no changes in the voltage-current properties after a cooling cycle. Figure 4\,A also depicts the temperature dependence of the sheet resistance of the Bi$_2$Te$_3$ films (10, 15, 20, 30 and 50\,QL). A clear metallic like behavior can be observed, with a saturation below 25\,K. The lowest attainable temperature of our system is about 13\,K, because no heat shielding is mounted at the sample position. There are no significant changes between the cool-down and warming up curves, demonstrating the excellent stability of the film and the contacts. ARPES data taken before and after the cooling cycle are identical, thus confirming spectroscopically the stability of the films.

Comparing the sheet resistances of the Bi$_2$Te$_3$ films with thicknesses ranging from 10 to 50\,QL (Fig. 4\,A), we notice that there is a scatter with a tendency to have lower resistances with thicker films. This variation is shown in Fig. 4\,B. We can clearly observe, however, that the resistance is not inversely proportional with the thickness; instead, it varies by a factor of only 1.3 at  14\,K and 1.5 at 295\,K, when going  from 10 to 50 QL. We can take this as evidence that the surface is indeed dominating the transport properties in our Bi$_2$Te$_3$ films. The corresponding ARPES data (Fig. S7) reveal slight variations of the position of the Fermi level within the gap. Together with the observation that the sheet resistance is practically constant for thickness $>$20-30\,QL, we deduce that possibly some minute amounts of Te deficiency may have occurred during the initial stages of the growth.

Figure 4\,B shows that the surface conductance $G_s$ converge to 40\,e$^2$/h and 25\,e$^2$/h for low and high temperatures, respectively. The surface charge carrier concentrations $n_s$ can be estimated from the corresponding ARPES data, using $n_s\sim k_F^2$ (Fig. S7). We obtained values in the range of 2-4$\cdot10^{12}$\,cm$^{-2}$, which correspond to average mobilities of about  3,000\,cm$^{2}$/Vs, with 4,600\,cm$^{2}$/Vs as highest value at 14\,K and 1,600\,cm$^{2}$/Vs at room temperature, see Fig. 4\,C. These mobilities  are higher than reported in the literature  so far for thin films \cite{Bansal2012,Taskin2012}.
We would like to remark that there are different methods to obtain surface carrier densities, either directly from the  Hall measurements \cite{Kim2012,Bansal2012} or through $k_F$ determined from SdH oscillations \cite{Butch2010,Qu2010} or  ARPES \cite{Aitani2013,Barreto2014,Zhu2011}. The SdH method is frequently used and gives good agreement on $k_F$ from ARPES \cite{Taskin2012}.
We note that SdH resistance oscillations show up for magnetic fields $B>\Phi_0k_F^2/2\pi^2\hat{G}_s$, where $\Phi_0=h/2e$ and $\hat{G}_s=G_s/(e^2/h)$. For typical $k_F=0.07\,$\AA$^{-1}$ and $\hat{G}_s=40$ we expect SdH oscillations to appear already above B\,$>$\,1.3\,T for our pristine high mobility films.

\begin{figure*}[tb]
\begin{center}
\centerline{\includegraphics[width=0.8\textwidth]{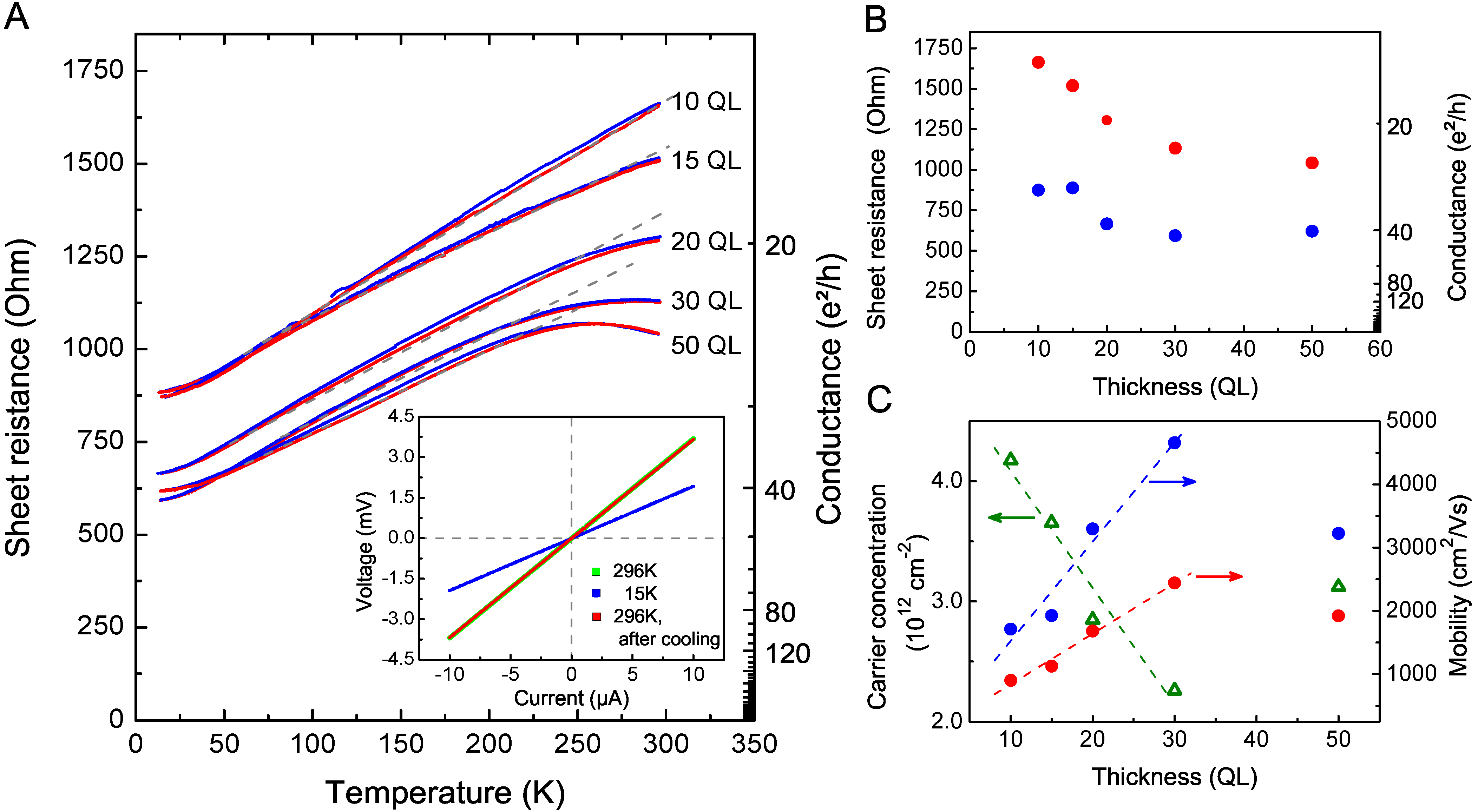}}
\caption{\textit{In-situ} transport properties of the Bi$_2$Te$_3$ thin films. {\bf A}, Temperature depended sheet resistance of the thin films ranging from 10\,QL to 50\,QL. The corresponding ARPES images can be found in Supporting Information S7. The cool-down (blue) and warm-up (red) curves are shown for each thickness.  The inset displays an exemplary I-V characteristic of a 10\,QL Bi$_2$Te$_3$ thin film. The linear behavior prove the ohmic contacts within the whole temperature range; green: before cool down at 296\,K, blue: 15\,K, red 296\,K after the thermal cycle.
{\bf B},  Variation of $R$ vs.\,thickness at low (blue dots) and room temperature (red dots).
{\bf C}, Charge carrier concentrations (green triangles) calculated from the ARPES spectra and resulting mobility values for the different film thicknesses at room temperature (red dots) and  14\,K (blue dots). The dashed lines are guides to the eye.}\label{pic4_transport}
\end{center}
\end{figure*}

For 3D TI thin films theory predicts a $R$ vs.\,$T$ behavior according to the electron-phonon scattering mechanism (quasiclassical Boltzmann transport theory) \cite{Giraud2012}, similar to graphene. The theory proposes a $\rho\propto T^4$ behavior at low temperatures and $\rho\propto T$ for high temperatures. The cross-over from the low to the high temperature regime is defined by the Bloch-Gr\"{u}neisen temperature $T_{BG}= 2\hbar k_Fc_S/k_B$. Here $c_S$ denotes an average speed of sound in the material ($\bar{c}_S^{Bi_2Te_3}=$2,200\,m/s) and $k_B$ is the Boltzmann constant. The Fermi wave vector $k_F$ is calculated for each sample from the measured ARPES spectra by using $\epsilon_F=\hbar v_F k_F$. The value of $\epsilon_F$ is determined from the energy distance of the chemical potential to the extrapolated Dirac Point, which for Bi$_2$Te$_3$ is hidden inside the bulk valence band. In this case we obtain a Bloch-Gr\"{u}neisen temperature in the range of  $T_{BG}\approx$22\,K for our films. Because the lowest attainable temperature of our experimental system is about 13\,K, the low temperature regime is not experimentally accessible. However, in the high temperature limit $T\gg T_{BG}$, we observe the expected linear $R$ vs.\,$T$ above $\sim$90\,K. However, for the thicker Bi$_2$Te$_3$ film (30 and 50\,QL), there is a deviation from the linear behavior above $\sim$230\,K. The underlying mechanism of the  decreasing resistivity could be attributed to the formation of thermally activated charge carriers.

\subsection{Effect of surface contamination}
Inside the UHV chamber with pressures in the low $10^{-10}$\,mbar range, the Bi$_2$Te$_3$ thin films show no detectable sign of aging in ARPES or in conductance, even after more than a week. Therefor the concentration of the typical residual gases (H$_2$, O$_2$, CO, CO$_2$ and H$_2$O) is too low to induce measurable degradation and/or that the surfaces of our films are quite inert. As for bulk samples, aging effects within hours after cleavage in UHV were reported \cite{Chen2009,Park2010,Zhou2012}.

To investigate the influence of surface contamination on the conductance of topological insulator surfaces, we exposed our Bi$_2$Te$_3$ samples to pure oxygen at a pressure of 1$\cdot10^{-6}$\,mbar for 10\,min. ARPES revealed no effect on the band structure or on the position of the chemical potential (Fig. S8). Consequently, pure oxygen can be excluded as a source of degradation of our surfaces. This finding is in contrast with earlier work \cite{Zhou2012} in which aging has been reported on adsorption of atoms or molecules. We therefore arrive at the conclusion that the inertness of our films must be related to the essentially perfect stoichiometry of the Te also at the surface.

We now study the influence of air in an attempt to assess the effect of conductivity experiments carried out \textit{ex-situ}. Fig. 5 depicts the sheet resistance and ARPES of a 10\,QL Bi$_2$Te$_3$ film, before (pristine) and after exposure to air. In this experiment the pristine sample was taken out to ambient atmosphere for 5\,min, and then introduced back into the load lock and pumped down to UHV conditions over night (12\,hours). No further treatments, like heating up or exposing to solvents, were done to the sample, which are usually required during contacting with Ag-paint or lithographic processes. We can clearly observe from Fig. 5\,A that the sheet resistance has dropped by about 200\,$\Omega$. ARPES reveals that the chemical potential has moved by about 50\,meV: whereas the pristine sample shows only the surface states intersecting the Fermi level, the air-exposed sample has its BCB shifted so much downwards that it becomes occupied (the bottom of the BCB is about 5\,meV below the Fermi level). Although all the band features remain intact, the conductance of the film is now clearly affected by the filling of the BCB.

\begin{figure*}[tb]
\begin{center}
\centerline{\includegraphics[width=0.8\textwidth]{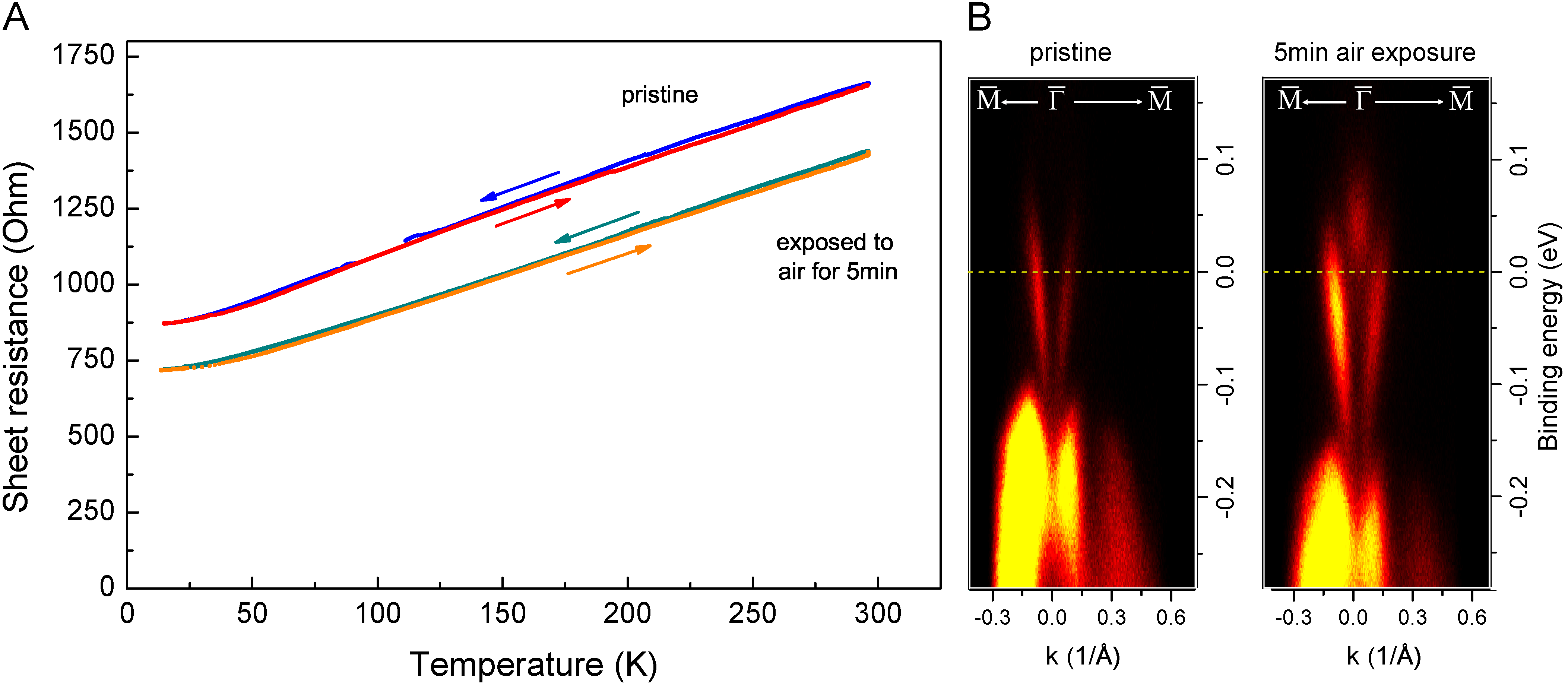}}
\caption{The effect of  air exposure on the electronic structure  and transport properties.  {\bf A}, $R$ vs. $T$ for a 10\,QL  Bi$_2$Te$_3$ thin film as grown (blue and red curve) and the same film after exposure to air for 5\,min and 12\,hours in vacuum (green and yellow curve). The air exposure causes a decrease in sheet resistance by 200\,$\Omega$.
{\bf B}, Corresponding ARPES spectra. The indirect band gap of the Bi$_2$Te$_3$ films can be determined to $E_{gap}\approx$145\,meV.}\label{pic5_air}
\end{center}
\end{figure*}

The charge carrier concentration for 2D Dirac materials follows $n_s(\epsilon_F)\propto\epsilon_F^2$, so the upward shift of the Fermi level by 50\,meV would lead to about 75\% more charge carriers in the sample. In addition, one has to take into account the carriers induced by the BCB occupation. Considering the formation of 2D quantum well states at the surface, due to band bending from  adsorbates \cite{Chen2012}, the 5\,meV occupation of the BCB can be estimated in a simple model to cause in total 83\% more charge carriers, i.e. several 10$^{12}$cm$^{-2}$, than on the pristine Bi$_2$Te$_3$ surface. To quantify how these extra charge carriers will affect the film resistance is actually not straightforward, because one also has to consider the reduction of the surface carrier mobility due to scattering at the adsorbates \cite{Culcer2010}. Nevertheless, the experiment indicates that the overall effect of air exposure is to reduce the resistance. Although no traces of water could be detected in XPS (the concentration is several 10$^{12}$cm$^{-2}$, i.e. less than 1\% of a monolayer coverage), this is the most likely candidate to cause the band bending due to its polar character, since oxygen (see above) or nitrogen (too inert) can be excluded.

We note that our air-exposure experiment may not reveal the full effect of the contamination occurring during \textit{ex-situ} transport measurements. Our exposure time is only 5\,min, and we allowed the sample to recover for 12\,hours in ultra-high vacuum conditions. The areal density of contaminants left adsorbed (several 10$^{12}$\,cm$^{-2}$) is therefore only a fraction of the full coverage possible (10$^{14}\sim\,$10$^{15}$\,cm$^{-2}$) under ambient conditions. The sheet resistance reduction in \textit{ex-situ} experiments can then be very much larger than the reduction from 800\,$\Omega$ to 600\,$\Omega$ that we have recorded. We would like to add that the extreme sensitivity of the resistance and the Fermi level position to minute amounts of external doping demonstrates once again that our films have extremely low charge carrier concentrations in the bulk.

A side effect of this air-exposure experiment is that now we can also determine the indirect band gap of our Bi$_2$Te$_3$ thin films. The ARPES spectrum in Fig. 5\,B displays that the energy separation between the BVB and BCB is $\sim$145\,meV, which agrees well with  reported experimental and theoretical results \cite{Chen2009,Zhang2009a}.

\subsection{Conclusion}
We  revealed the intrinsic surface transport in pristine thin films of the topological insulator Bi$_2$Te$_3$ with exceptionally high mobility. This progress has been achieved by all \textit{in-situ} experiments using a combination of thickness depended ARPES  and resistivity measurements. ARPES proved the position of the Fermi level inside the band gap prior and after the transport measurements, which always showed metallic like behavior of the TI thin films.  The degradation  effect of adsorbents on the surface of topological insulators, observed after exposure to ambient atmosphere, emphasizes the importance for carrying out these type of experiments under ultra-high vacuum conditions.  Alternatively, there is a need to find appropriate capping layers, which can preserve the Dirac surface states and the chemical potential so that their unique topological properties can be revealed by contact measurements in devices under ambient conditions.

\section{Experimental}
Epitaxial thin films of Bi$_2$Te$_3$ were grown on insulating, well lattice matched  epi-polished BaF$_2\,(111)$ substrates (mismatch $<$0.1\%, size 10$\times$10\,mm) and less-well matched Al$_2$O$_3$\,(0001) (mismatch 8.7\%) substrates (Fig. S9) by molecular beam epitaxy. High-purity (99.9999\%) elemental Bi and Te were evaporated from standard Knudsen cells, with flux rates of 1\,{\AA}/min for Bi and about 8\,{\AA}/min for Te; the fluxes were measured with a quartz crystal microbalance.
RHEED oscillations reveal a layer-by-layer growth mode with a typical growth rate of 0.3\,QL/min.
High-resolution ARPES measurements were performed using a non-monochromatized He discharge lamp with 21.2\,eV photon energy (He\,I line); XPS was done using a monochromatized Al-$K_{\alpha}$~ source.  All photoemission spectra were recorded with a VG~Scienta electron analyzer R\,3000  at room temperature. The spectra were calibrated by measuring the Fermi edge of a clean polycrystalline Ag sample. Temperature depended resistivity measurements were performed with a self-designed four-point probe setup mounted to a Janis~ST-400 continuous flow cryostat, allowing temperature changes from 300\,K to 13\,K. The probes were placed in a row with 2\,mm spacing between each contact. The sheet resistance of the films was determined using the equation $R_{sheet}= \pi/ln2 \cdot V/I$. Data acquisition was done by a Lock-In Amplifier (Zurich Instruments HF2LI) in combination with a home-made current source at 19\,Hz and 10\,$\mu$A bias current. Bi-directional I-V characterization was conducted in DC mode with the same current source and a Keithley 2000 digital multimeter. The preparation and all measurements were carried out \textit{in-situ} in a UHV system with a base pressure  in the low $10^{-10}$\,mbar range. Additional \textit{ex-situ} XRD characterization was performed with a PANanlytical X'Pert PRO diffractometer, using monochromatic Cu- $K_{\alpha}$ ($\lambda$= 1.54056\,\AA) radiation.

\begin{acknowledgments}
 We thank Dr. Steffen Wirth for very helpful discussions. J.S. is supported by the Max Planck -- UBC centre for Quantum Materials.
\end{acknowledgments}

%\bibliography{references}
\input{Hoefer_Intrinsic_surface_conduction_through_topological_surface_states_of_insulating_Bi2Te3_epitaxial_thin_films.bbl}

\clearpage

\pagenumbering{roman}
\renewcommand{\figurename}{FIG.\,\,S$\!\!$}
\makeatother
\setcounter{figure}{0}
\setcounter{page}{1}

\begin{figure*}
\begin{flushleft}
\Large{\textsf{\textbf{Supporting Information for: Intrinsic conduction through topological surface states of insulating Bi$_2$Te$_3$ epitaxial thin films}}} \\\bigskip
\indent \normalsize{\textsf{K. Hoefer et al. (2014)}}
\end{flushleft}
\end{figure*}

\begin{figure*}[htb]
\centering
\includegraphics[width=1\textwidth]{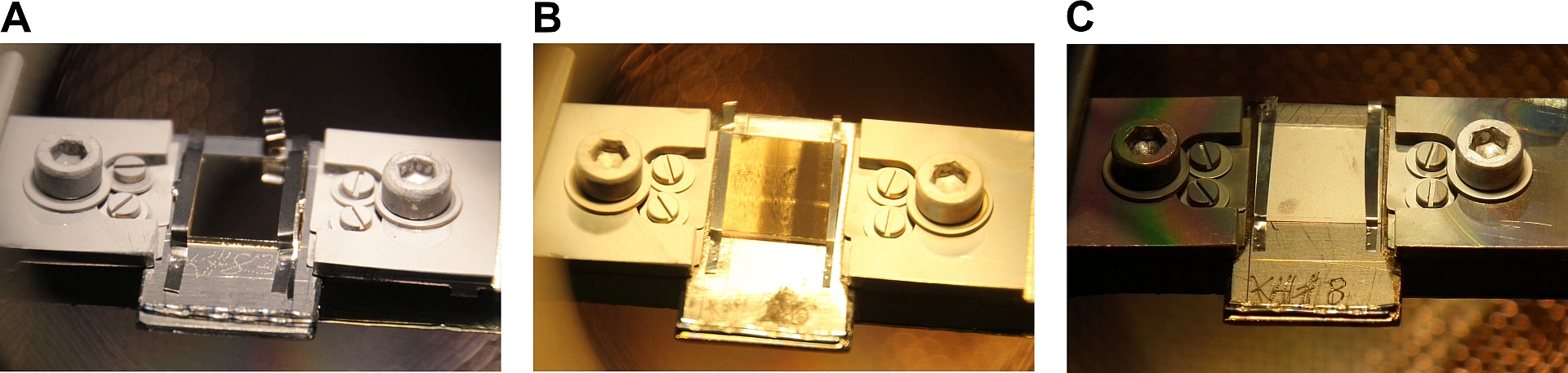}
\caption {Determination of the optimum substrate temperature.
{\bf A}, Sample grown at the optimum substrate temperature of 250\,$^\circ$C, which ensures full re-evaporation of excess Te, excellent surface mobility of the adatoms for layer-by-layer growth and a sticking coefficient close to 1 for Bi$_2$Te$_3$ and Bi. The samples are blackish and mirror-like.
{\bf B}, At 270\,$^\circ$C, the sticking coefficient of Bi$_2$Te$_3$ is already drastically decreased, as seen by the only partially covered substrate.
{\bf C}, At 300\,$^\circ$C substrate temperature  no deposition of Bi$_2$Te$_3$ occurs.\newline
The lowest substrate temperature to re-evaporate Te was determined by studying  the RHEED pattern of a vacuum-annealed substrate under Te flux. At 300\,$^\circ$C substrate temperature, the pattern is unaffected by the Te flux. The substrate temperature was then subsequently lowered in steps of 5\,K, and only at 170\,$^\circ$C the pattern starts to disappear, indicating that the Te is no longer fully re-evaporated.}
\end{figure*}

\begin{figure*}[htb]
\centering
\includegraphics[width=0.8\textwidth]{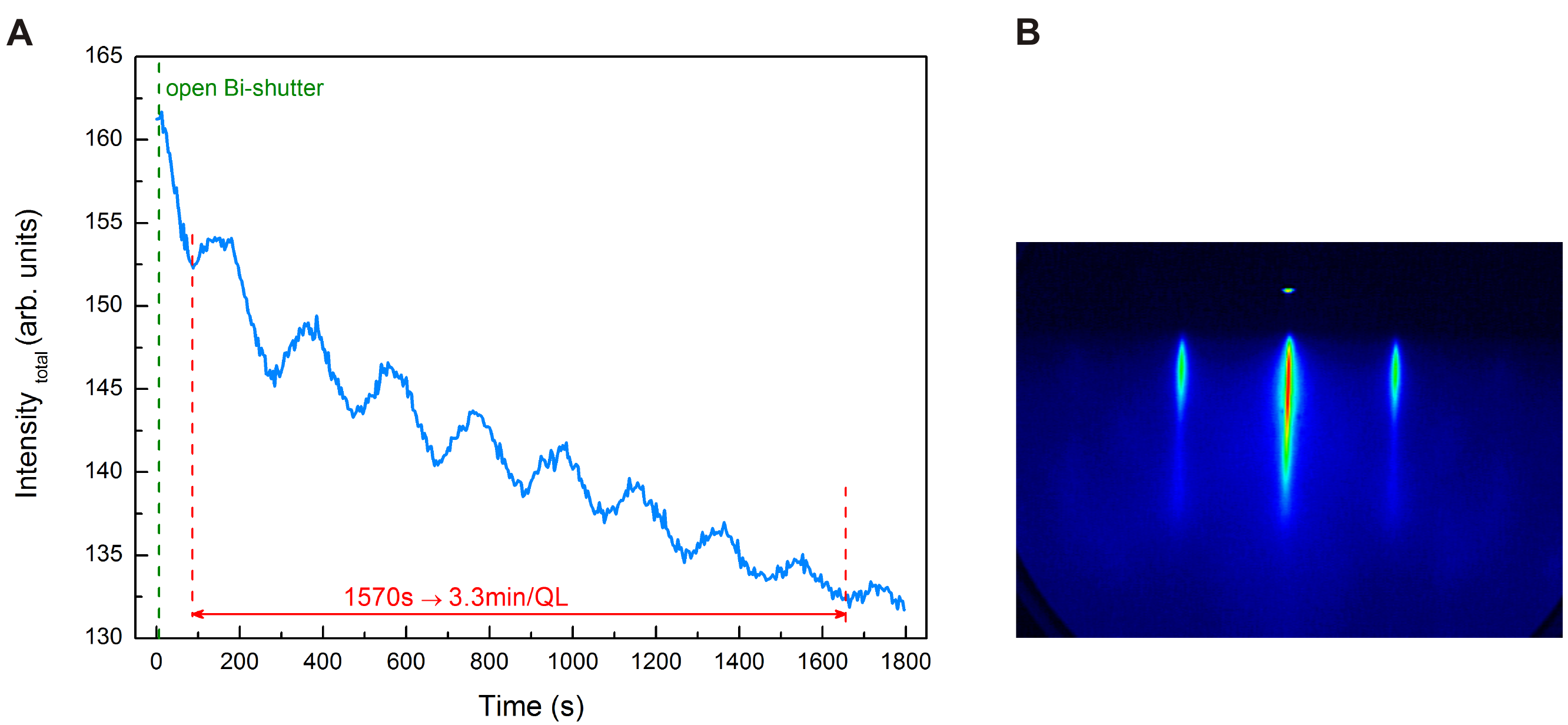}
\caption{\textit{In-situ} RHEED characterization during growth.
{\bf A}, Real time RHEED intensity oscillations of the specular spot showing a 2D layer-by-layer growth mode. The deposition rate is about 0.3\,QL/min, consistent  with the Bi flux of 1\,\AA/min.
{\bf B}, Intense streaky RHEED pattern of the 15\,QL thick film along the $[1\,\bar{2}\,1]$ direction depicts single crystalline film with a smooth surface.}
\end{figure*}

\begin{figure*}[ht]
\centering
\includegraphics[width=0.8\textwidth]{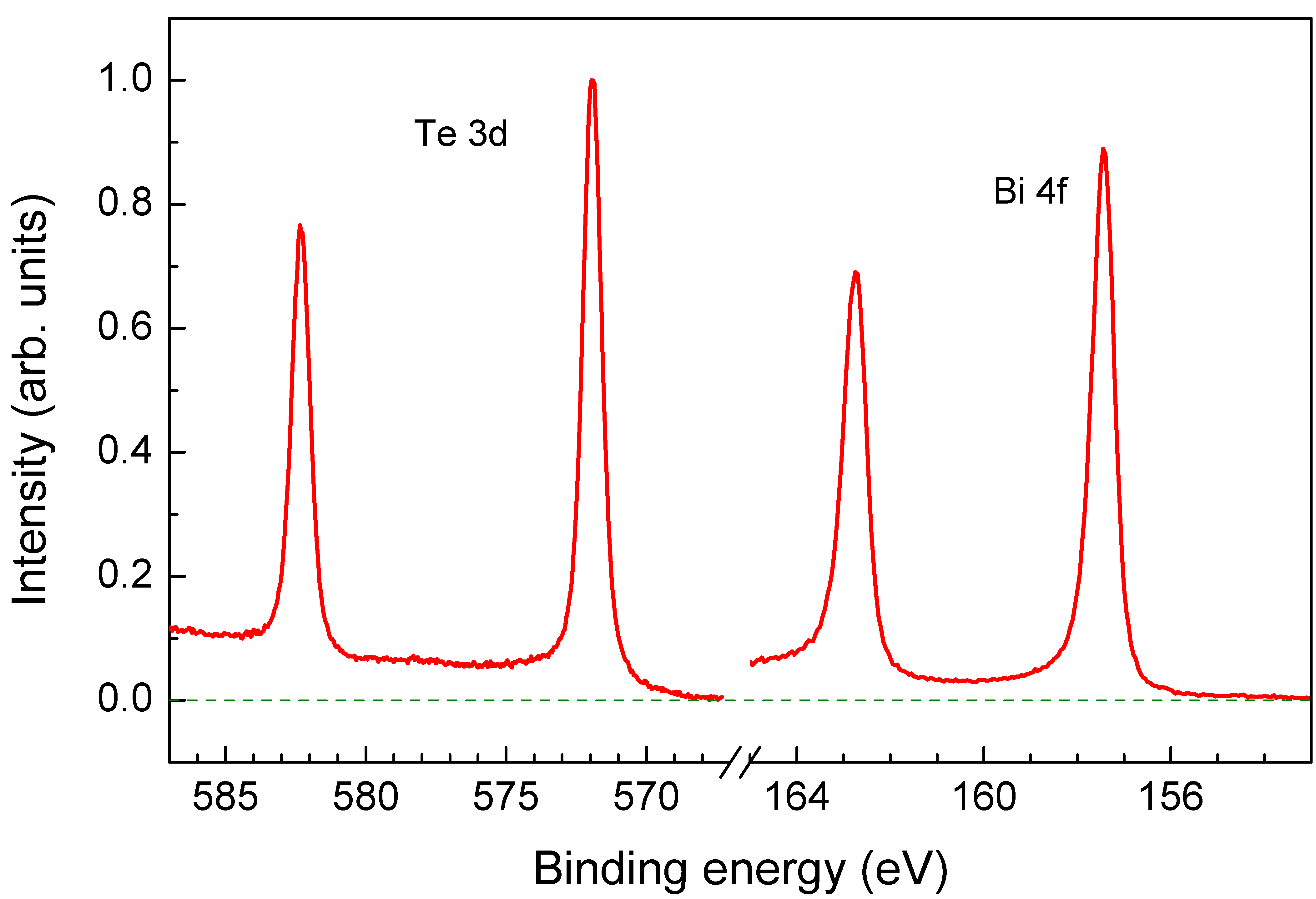}
\caption{X-ray photoelectron spectroscopy of the Te 3d and Bi 4f core levels. The very narrow and symmetric  core level lines indicate the absence of Te or Bi excess.}
\end{figure*}

\begin{figure*}[htb]
\centering
\includegraphics[width=0.8\textwidth]{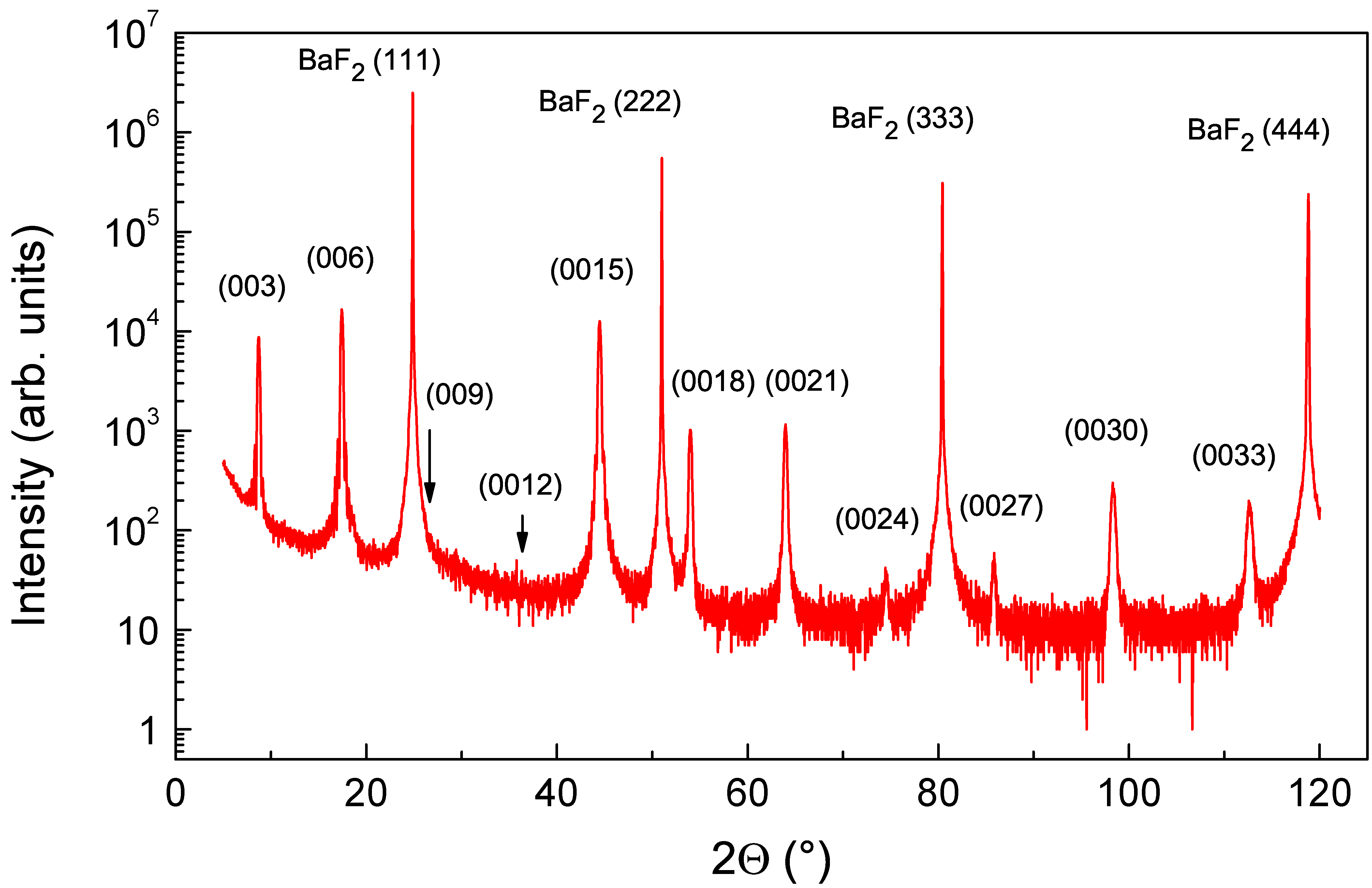}
\caption{\textit{Ex-situ} XRD  $\Theta-2\Theta-$ scan of a 30\,QL Bi$_2$Te$_3$ thin film shows only the 3n allowed reflection peaks, verifying the expected rhombohedral structure for this material system.}
\end{figure*}

\begin{figure*}[htb]
\centering
\includegraphics[width=0.8\textwidth]{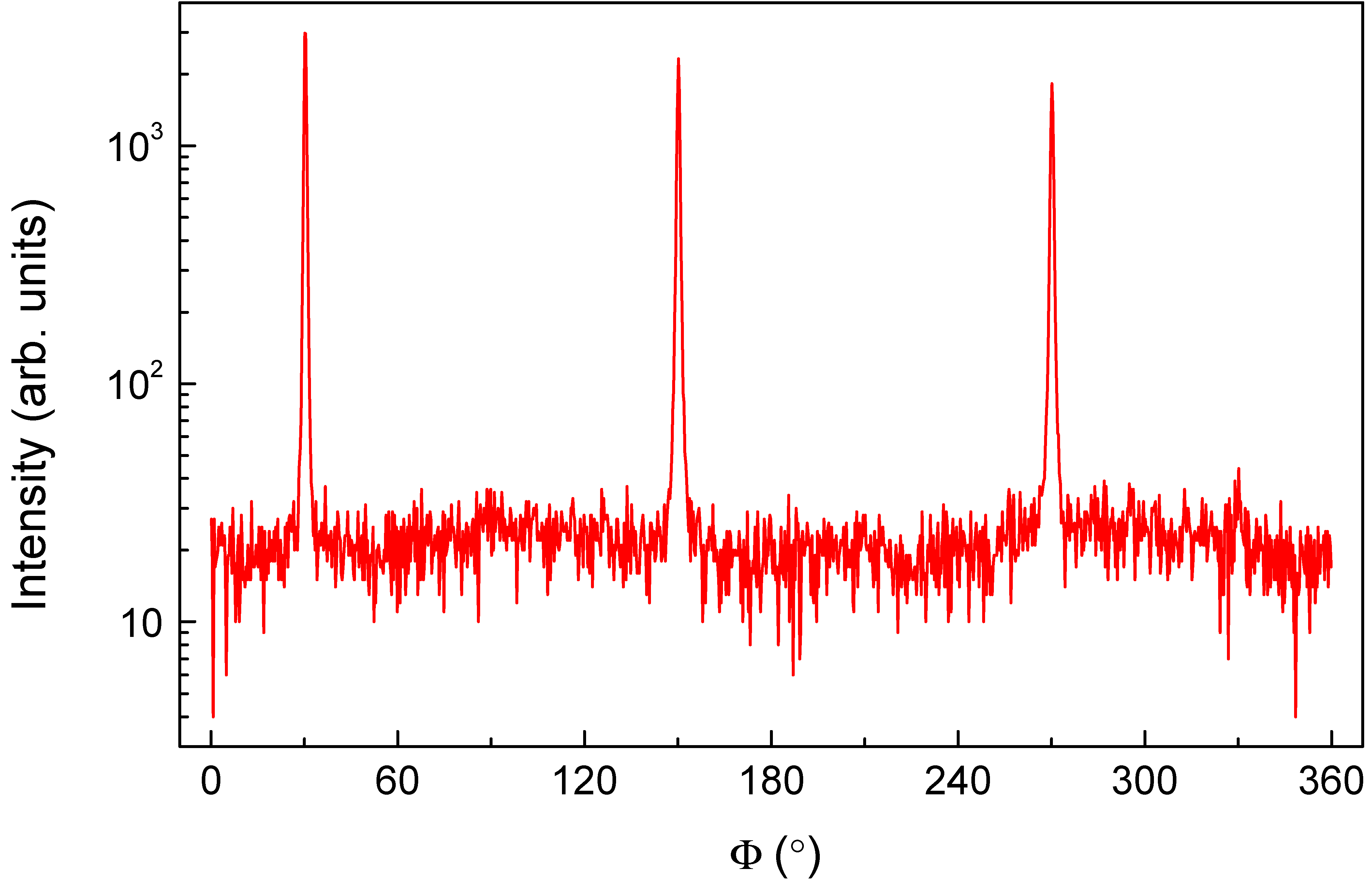}
\caption{XRD In-plane $\Phi-$rotation scan around the $(0\,1\,5)$ Bi$_2$Te$_3$ peak. The reflections occur every 120$^\circ$, further confirming the 3-fold symmetry observed in LEED and ARPES. These measurements prove the single-domain nature of our films.}
\end{figure*}

\begin{figure*}[htb]
\centering
\includegraphics[width=0.8\textwidth]{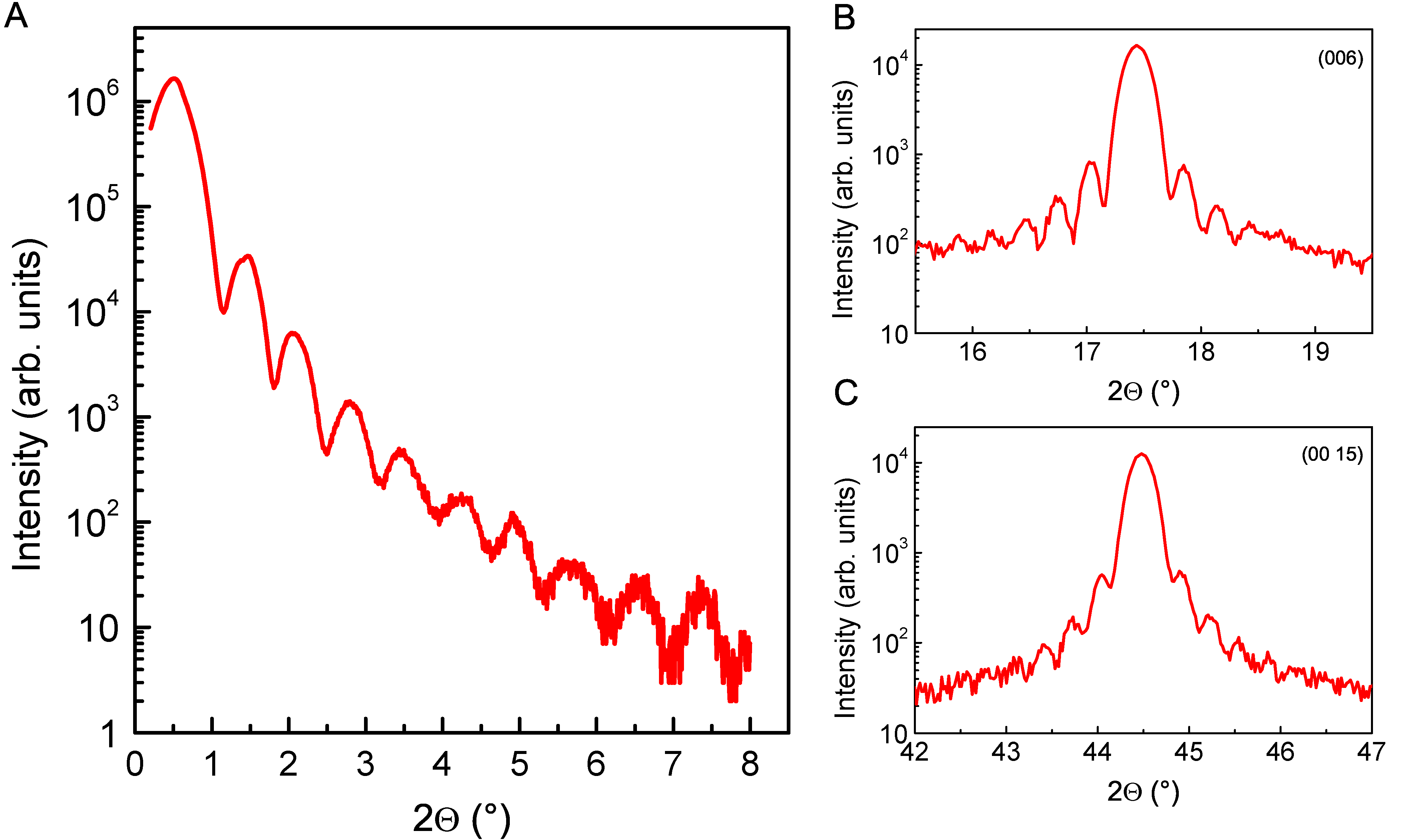}
\caption{\textit{Ex-situ} X-ray diffraction.
{\bf A}, X-ray reflectivity (XRR) shows distinct Kiessig fringes over more than 5 orders of magnitude in intensity and up to $2\Theta=$\,8\,$^\circ$, indicating a very low long range surface roughness of less than 0.2\,nm of the 12\,QL Bi$_2$Te$_3$ films.
{\bf B} and {\bf C} display a zoom-in of the  $(0\,0\,6)$ and $(0\,0\,15)$ Bragg peak of a 30\,QL thin film (Fig. S4), respectively. The observed fringes persist even at the $(0\,0\,15)$ peak at $2\Theta=$\,44.5\,$^\circ$, supporting the flat surface morphology.}
\end{figure*}

\begin{figure*}[htb]
\centering
\includegraphics[width=1\textwidth]{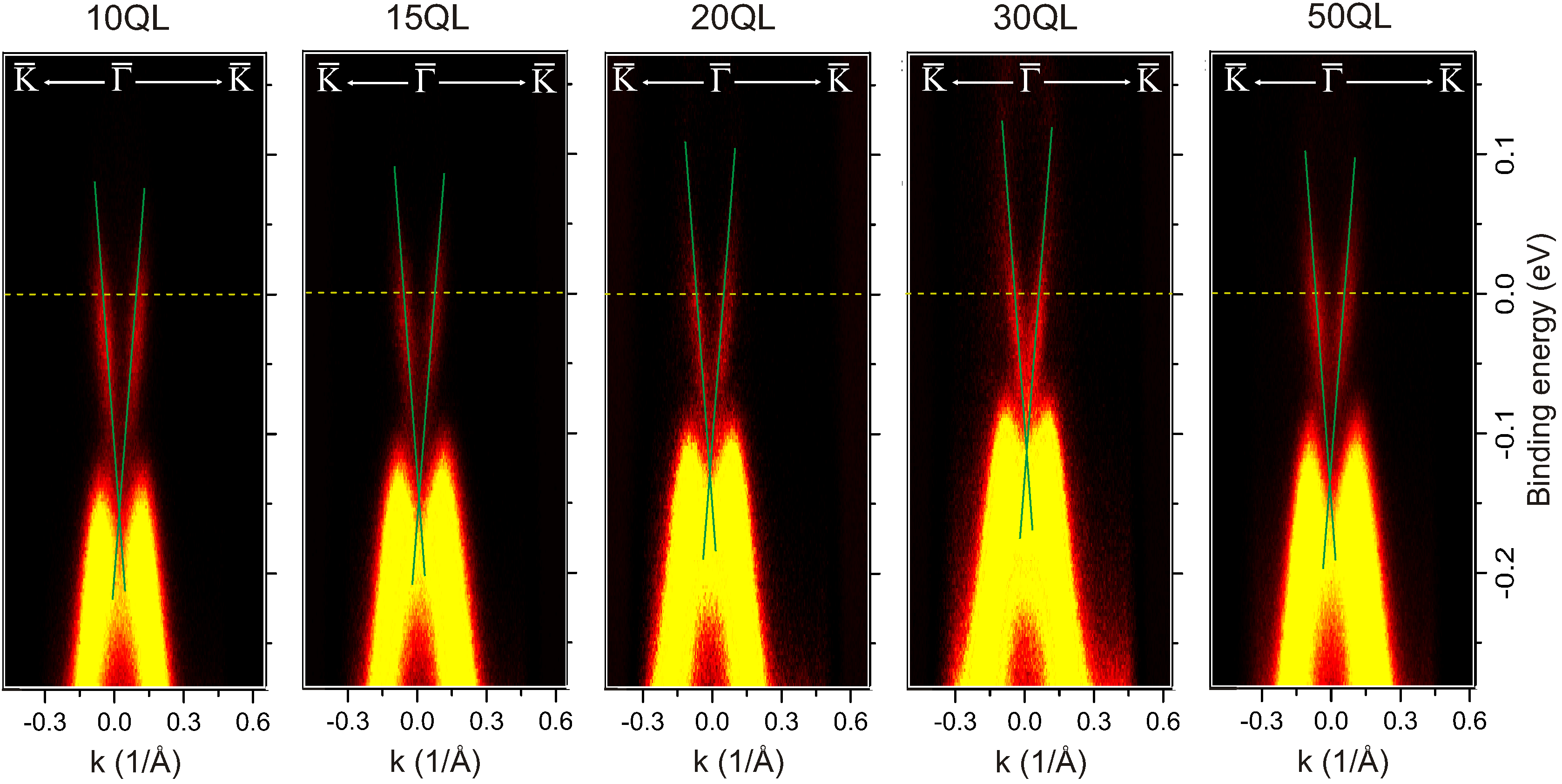}
\caption{ARPES images of the samples used for the thickness depended resistivity measurements, ranging from 10 to 50\,QL. For all samples only the SS are intersecting the Fermi level. We determined the following energy distance of the Fermi level  to the extrapolated Dirac point: 155\,meV for 10\,QL, 145\,meV for 15QL, 128\,meV for 20\,QL, 114\,meV for 30\,QL and 134\,meV for 50\,QL.}
\end{figure*}

\begin{figure*}[htb]
\centering
\includegraphics[width=0.7\textwidth]{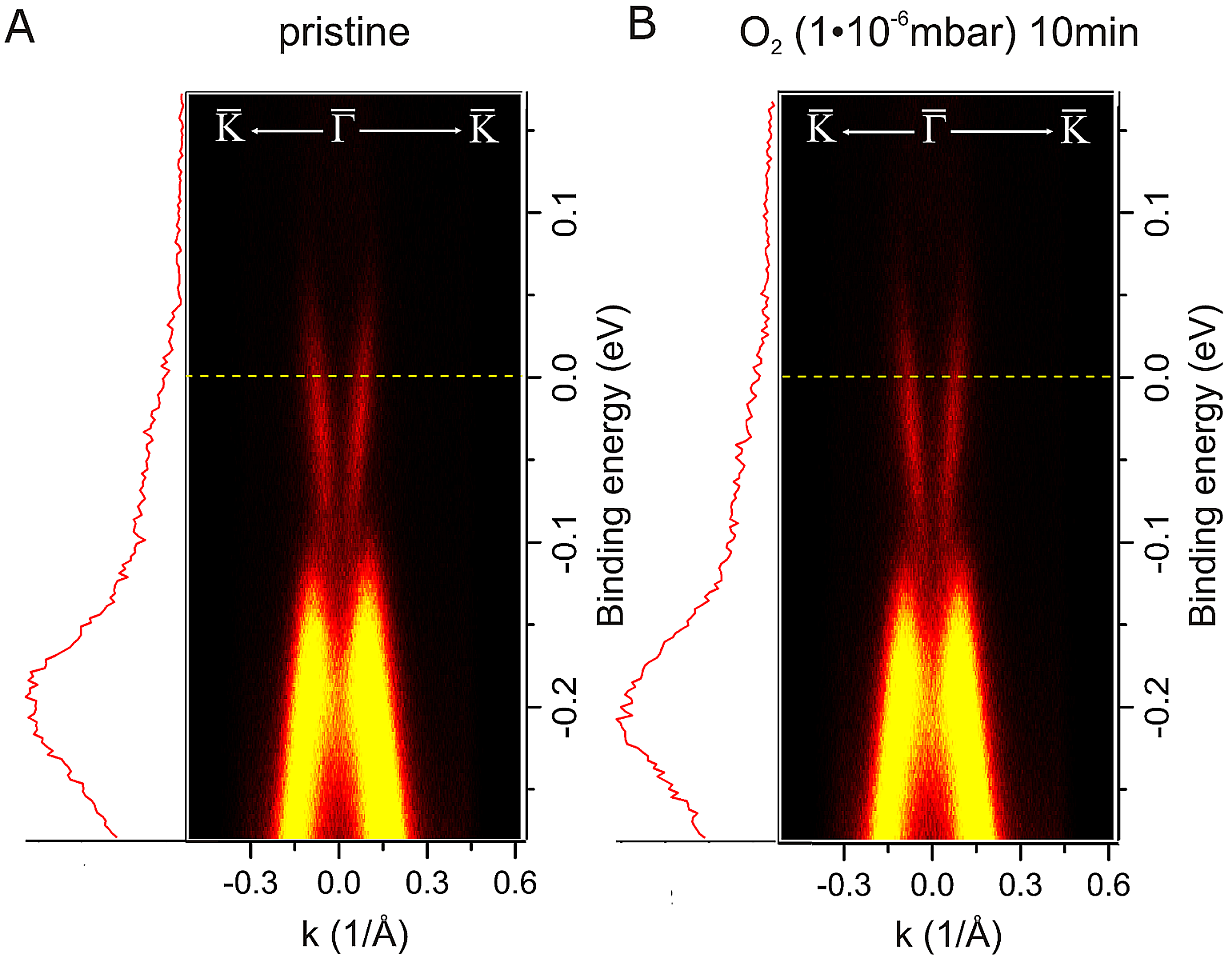}
\caption{Effect of Oxygen exposure.
{\bf A}, ARPES and image line profile at the  $\bar{\Gamma}-$point of the pristine surface.
{\bf B}, After exposure  to $1\cdot10^{-6}$\,mbar pure oxygen for 10\,min. The valence band structure and especially the Dirac cone are not affected. This observation indicates an essentially perfect Te stoichiometry on the surface, which is inert towards oxygen.}
\end{figure*}

\begin{figure*}[htb]
\centering
\includegraphics[width=0.26\textwidth]{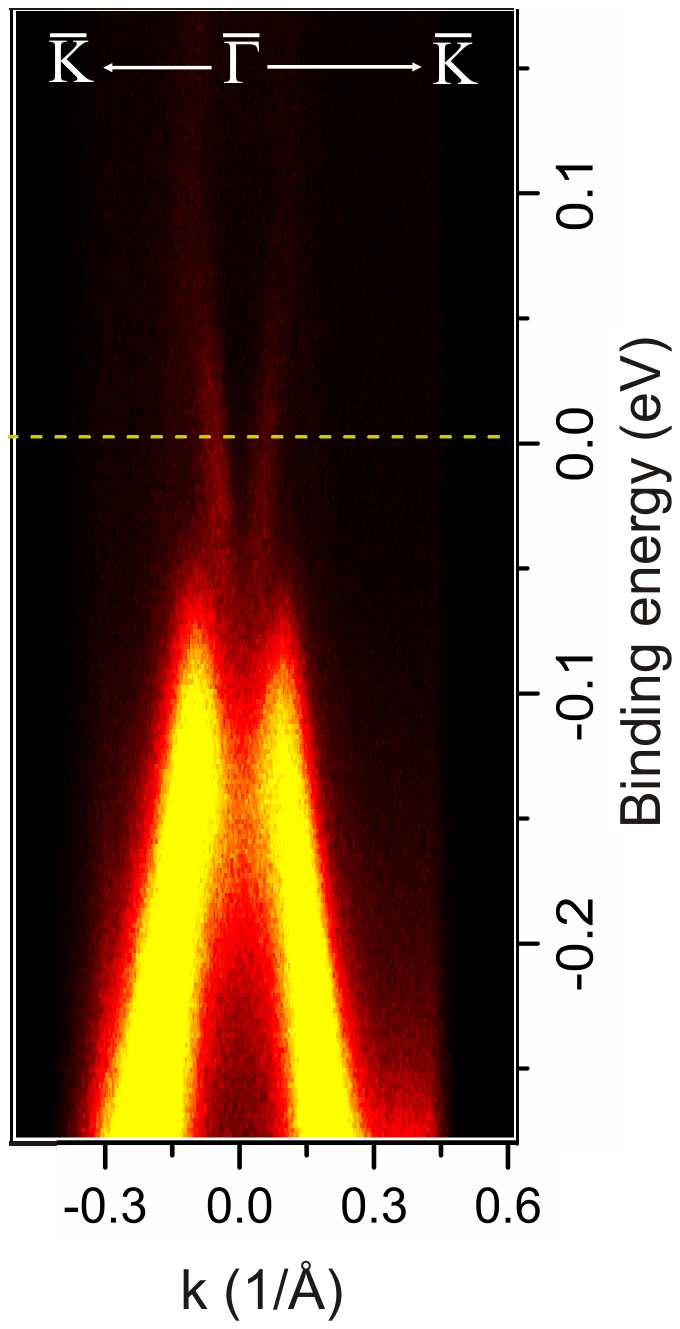}
\caption{ARPES image  of a 21\,QL Bi$_2$Te$_3$ film grown on Al$_2$O$_3$ (0001). Also for this sample only the SS are intersecting the Fermi level.}
\end{figure*}

\end{document}